\documentclass{PoS}
\usepackage{subfigure}  
\newcommand{\HBlabel}{Bechtle:2008jh,Bechtle:2011sb,Bechtle:2013gu,Bechtle:2013wla}
\newcommand{\FHlabel}{Heinemeyer:1998np,Heinemeyer:1998yj,Degrassi:2002fi,Heinemeyer:2007aq,Hahn:2013ria}
\def\cp{CP}
\def\gev{\,\textrm{GeV}}
\newcommand{\mhp}{M_{H^{\pm}}}
\newcommand{\hpm}{H^{\pm}}
\newcommand{\Zb}{\hat{\textbf{Z}}}
\newcommand{\pat}{\phi_{A_t}}
\newcommand{\mhmod}{M_h^{\rm{mod+}}}
\newcommand{\bb}{b\bar{b}}

\title{Interference effects in MSSM Higgs searches at the LHC}

\ShortTitle{MSSM Higgs Interference}

\author{\speaker{Elina Fuchs}\\
        Weizmann Institute of Science, 760001 Rehovot, Israel\\
        E-mail: \email{elina.fuchs@weizmann.ac.il}}

\author{Georg Weiglein\\
       Deutsches Elektronen-Synchrotron (DESY), Notkestr. 85, 22607 Hamburg, Germany\\
       E-mail: \email{georg.weiglein@desy.de}}

\abstract{
Complex parameters in the MSSM lead to mixing and interference between the two heavier neutral CP-even and CP-odd Higgs states.
These effects can become very large in the case of almost degenerate states.
In a CP-violating benchmark scenario, we investigate phenomenological implications of such interferences in view of the LHC searches for heavy Higgs bosons decaying to a pair of $\tau$-leptons and produced in gluon fusion and in association with $b$-quarks. Strongly destructive effects leave parameter regions unconstrained that would be regarded as excluded if no interference terms were taken into account.\\

}

\FullConference{38th International Conference on High Energy Physics\\
		3-10 August 2016\\
		Chicago, USA}

\begin{document}
\section{Introduction}
Searches for additional Higgs bosons
have so far been interpreted within models beyond the Standard Model (BSM) by assuming the on-shell production of an unstable scalar and its subsequent decay. For quasi mass-degenerate particles that can appear as intermediate states between a given initial and final state, however, such a single-resonance approach or the incoherent sum of two resonance contributions does not necessarily hold. If the mass difference is smaller than the sum of their total widths, the two resonances overlap. This can lead to a potentially large interference term, which is neglected in the standard narrow-width approximation (NWA), but can be taken into account in the full calculation or in a generalised NWA\,\cite{Fuchs:2014ola}.

An important example for such a case are the two heavier neutral Higgs bosons of the Minimal Supersymmetric Standard Model (MSSM) which are almost mass-degenerate in the decoupling limit. 
In the presence of complex MSSM parameters, they mix in a CP-violating way and interfere with each other.
For heavy Higgs production in gluon fusion and in association with $\bb$ and the decay into a pair of $\tau$-leptons, we study the impact of such mixing and interference effects on the exclusion bounds within the MSSM with complex parameters (see also~\cite{Fuchs:2015jwa,Fuchs:2017wkq}).

\section{MSSM Higgs sector with complex parameters}
Among the three neutral MSSM Higgs bosons, there are at lowest order the $\cp$-even states $h$ and $H$ as well as the $\cp$-odd $A$. The charged Higgs bosons are denoted by $\hpm$. While the Higgs sector is $\cp$-conserving at lowest order, non-vanishing imaginary parts of MSSM parameters enter the Higgs sector via loop corrections, leading to $\cp$-violating mixing of the neutral $\cp$-eigenstates $i=h,H,A$ into the mass eigenstates $h_1,h_2,h_3$ with masses $M_{h_a}$ and total widths $\Gamma_{h_a},\,a=1,2,3$. The corresponding mixing factors are given by the on-shell wave function normalisation factors $\Zb_{ai}$ evaluated at the complex poles $\mathcal{M}^{2}_{h_a}=M_{h_a}^{2}-iM_{h_a}\Gamma_{h_a}$ which can be obtained from \texttt{FeynHiggs}~\cite{\FHlabel}. We adopt the renormalisation scheme of Ref.\,\cite{Frank:2006yh}.

The phases of MSSM parameters such as the trilinear couplings $A_f$ of the sfermions $\tilde{f}$ and the mass parameters of gauginos, $M_1,M_3$, and higgsinos, $\mu$,  are in particular bounded by constraints from electric dipole moments (see e.g. Refs.\,\cite{Barger:2001nu,Li:2010ax,Engel:2013lsa}). However, constraints on the phases $\phi_{A_{f_3}}$ of the trilinear couplings of the third generation are weaker than for the first two generations. In addition, $\pat$ has the largest impact on the Higgs sector. Therefore we focus on the consequences of this phase for the mixing and interference effects and set $\pat=\phi_{A_b}=\phi_{A_{\tau}}$ in this work. The impact of the variation of the gluino phase will be discussed in~\cite{IntCalc:InProgress}.

\section{$\cp$-violating interference of MSSM Higgs bosons}
\begin{figure}[ht!]
 \begin{center}
  \subfigure[$b\bar{b}\rightarrow h_a\rightarrow \tau^{+}\tau^{-}$.]{\includegraphics[width=0.495\textwidth]{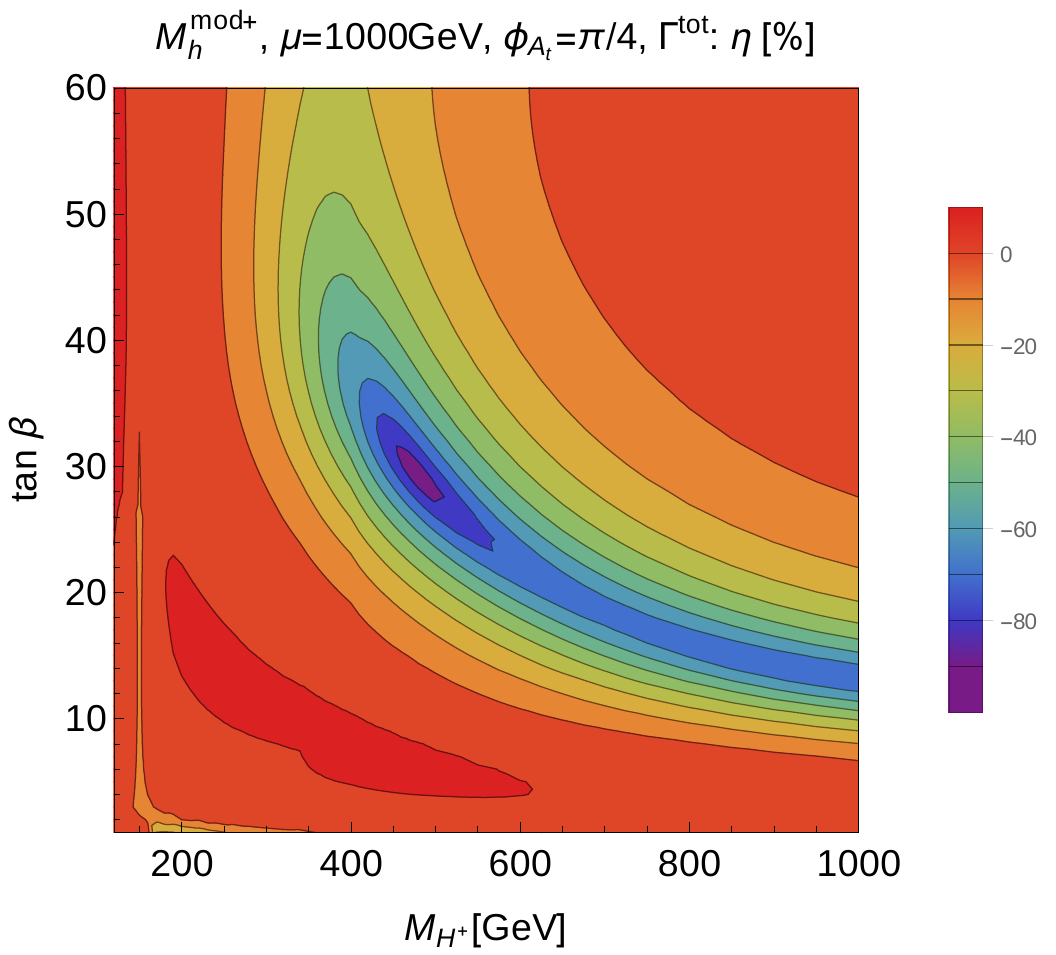}\label{fig:etabb}}
  \subfigure[$gg\rightarrow h_a \rightarrow \tau^{+}\tau^{-}$.]{\includegraphics[width=0.495\textwidth]{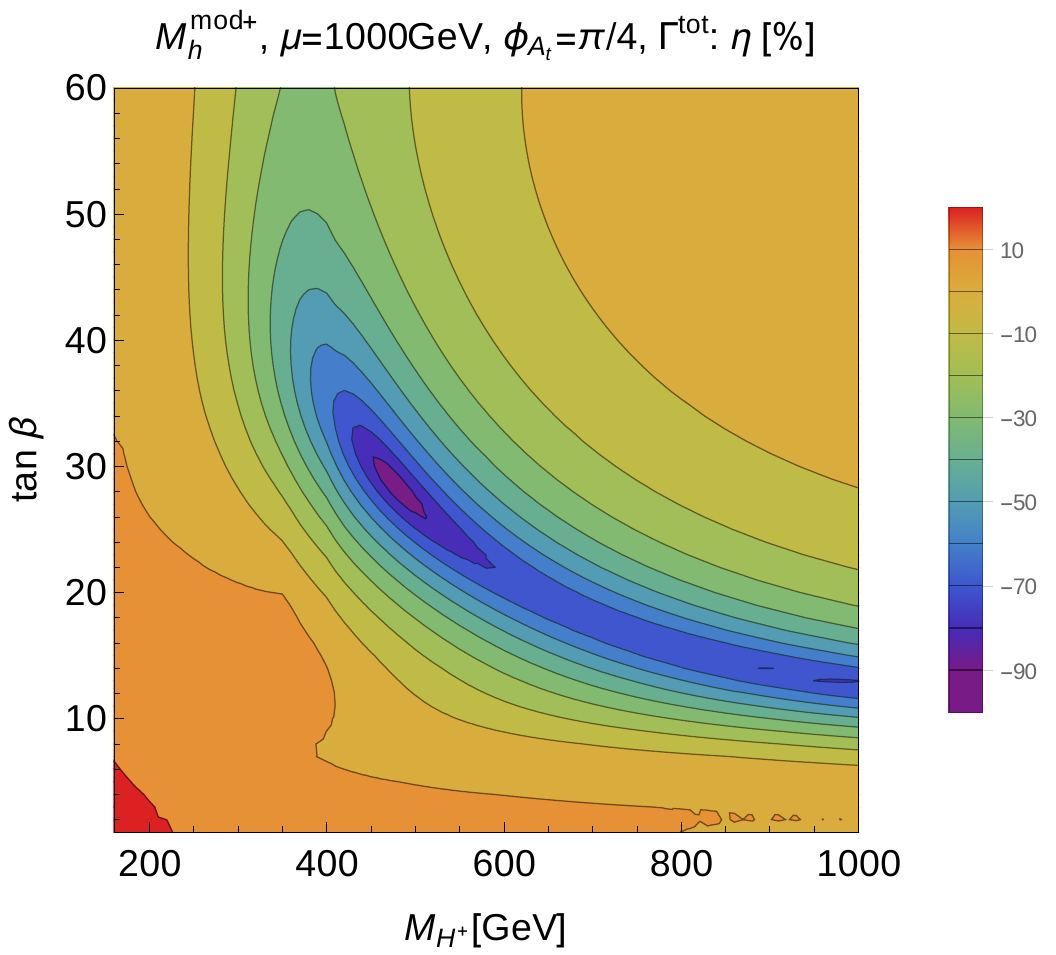}\label{fig:etagg}}
  \caption{Relative interference contribution $\eta[\%]$ of the Higgs bosons $h_1,h_2,h_3$ decaying to $\tau^{+}\tau^{-}$ in the complex $\mhmod$-scenario with $\mu=1000\gev$ and $\pat=\frac{\pi}{4}$. (a) $b\bar{b}$ initial state, (b) $gg$ initial state (note the different scale of the colour code).}
 \end{center}
 \label{fig:eta}
\end{figure}

While for low and medium values of $\tan\beta$ the production of neutral MSSM Higgs bosons at the LHC is dominated by gluon fusion, $gg\rightarrow h_a\,(a=1,2,3)$, at high $\tan\beta$ the associated production with $b\bar{b}$ dominates. Regarding the decay, the $\tau^{+}\tau^{-}$-\,channel provides the strongest constraints\,\cite{CMS:2015mca,Aad:2014vgg,CMS:2016pkt,Aaboud:2016cre} for intermediate and large $\tan\beta$. For interference effects at low $\tan\beta$ in the $t\bar t$ final state, see~\cite{Bernreuther:2015fts,Carena:2016npr}. We calculate the partonic cross sections of 
(i) $b\bar{b}	~\rightarrow h_a~\rightarrow~\tau^{+}\tau^{-}$,
(ii) $gg	~\rightarrow h_a~\rightarrow~\tau^{+}\tau^{-}$.
In our approach, we combine finally the existing precise predictions for separate Higgs production and decay (here from \texttt{FeynHiggs-2.10.2} \cite{\FHlabel}) with the relative interference contribution. We compute the interference term in the $2\rightarrow 2$ processes without NWA, including propagator corrections, but no vertex and real corrections, which factorise and will be already accounted for in the production cross sections and branching ratios. 
Hence we calculate both process at leading order, i.e.~(i), which is part of the $b\bar{b}h_a$-associated production, at tree-level, and (ii) with one-loop diagrams of the first vertex, where the $t,\tilde t, b, \tilde b$ loops dominate. 
The propagator-type corrections are accounted for by using Higgs masses, total widths and $\Zb$-factors from \texttt{FeynHiggs} at 2-loop order. The Higgs resonances are parametrised as Breit-Wigner propagators, and the mixing is accounted for by $\Zb$-factors, which yields a good approximation of the full propagators\,\cite{Fuchs:2015jwa,Fuchs:2016swt}. 
The cross section of the complete process contains the \textit{coherent} sum of the amplitudes with $h_1,h_2,h_3$-exchange in the $s$-channel, 
i.~e. $\sigma_{\rm{coh}} = \sigma\left(|h_1+h_2+h_3|^{2}\right)$, including the interference term, which is neglected in the \textit{incoherent} sum,
$\sigma_{\rm{incoh}} = \sigma\left(|h_1|^{2}+|h_2|^{2}+|h_3|^{2}\right)$. Consequently, 
$ \sigma_{\rm{int}} = \sigma_{\rm{coh}}-\sigma_{\rm{incoh}}$, and we define the relative interference contribution
\begin{eqnarray}
 \eta :=\frac{\sigma_{\rm{int}}(\pat)}{\sigma_{\rm{incoh}}(\pat)}\label{eq:eta}.
\end{eqnarray}
In the case of $\pat=0$, only $h$ and $H$ can interfere whereas a non-vanishing value of $\pat$ induces a $3\times 3$ mixing and the interference between all of the neutral Higgs bosons $h_1,h_2,h_3$.
For the numerical evaluation, we define a $\cp$-violating benchmark scenario as a modification of the standard $\mhmod$-scenario\,\cite{Carena:2013ytb} by setting $\mu=1000\gev$ (as also proposed in \cite{Carena:2013ytb}) and introducing the non-vanishing phase $\pat=\frac{\pi}{4}$. 
In this $\cp$-violating scenario, Fig.\,\ref{fig:eta} shows the relative interference effect $\eta$ in the $\mhp-\tan\beta$ plane, Fig.\,\ref{fig:etabb} for the $b\bar{b}$-initiated process (i), and Fig.\,\ref{fig:etagg} for gluon fusion (ii), both with $\tau^{+}\tau^{-}$ in the final state.
Due to the approximate degeneracy of $M_{h_2}$ and $M_{h_3}$ in the decoupling limit and the significant mixing of $H,A$ into $h_2,h_3$ in sizeable parts of the parameter space, the interference term becomes very large in both processes. Around $\mhp=480\gev,\,\tan\beta=29$, the relative interference contribution ranges down to $\eta\simeq-97\%$, surrounded by a considerable, destructive interference ``valley''.

\section{Impact of interference terms on exclusion limits}
The significant destructive interference terms suppress the predicted cross sections and need to be included in a consistent evaluation of exclusion bounds. For the comparison of the model prediction and LHC data within \texttt{HiggsBounds}\,\cite{\HBlabel} we use the following approximate treatment. The overall interference term per process is split into the three combinations of two interfering Higgs bosons, respectively,
$
 \sigma_{\rm{int}} = \sigma_{\rm{int}_{12}} + \sigma_{\rm{int}_{13}} +\sigma_{\rm{int}_{23}}\,.
 \label{eq:intsplit}
$
The individual cross sections $\sigma_{h_a}$
of each of the Higgs bosons $h_a$ are modified by the relative interference contribution
\begin{eqnarray}
  \eta^{P}_a = \frac{\sigma^{P}_{\rm{int}_{ab}}}{\sigma_{h_a}^{P}+\sigma_{h_b}^{P}}
   +\frac{\sigma^{P}_{\rm{int}_{ac}}}{\sigma_{h_a}^{P}+\sigma_{h_c}^{P}},\label{eq:eta_a}
\end{eqnarray}
where $a,b,c=1,2,3$, and $P=gg,b\bar b$ denotes the production mode.
The relative interference contributions $\eta_a^P$ are used to rescale the ratio of the production cross sections $P\rightarrow h_a$ in the MSSM with respect to the SM as input for \texttt{HiggsBounds} in the following way,
\begin{eqnarray}
 \frac{\sigma^{\rm{MSSM}}(P\rightarrow h_a)}{\sigma^{\rm{SM}}(P\rightarrow h_a)}
~~ \longrightarrow ~~
 \frac{\sigma^{\rm{MSSM}}(P\rightarrow h_a)}{\sigma^{\rm{SM}}(P\rightarrow h_a)}\,\cdot (1+\eta_a^P),
\end{eqnarray}
leaving the branching ratios $\textrm{BR}(h_a\rightarrow \tau^{+}\tau^{-})$ unchanged. 

\begin{figure}[t]
 \begin{center}
\includegraphics[width=0.495\textwidth]{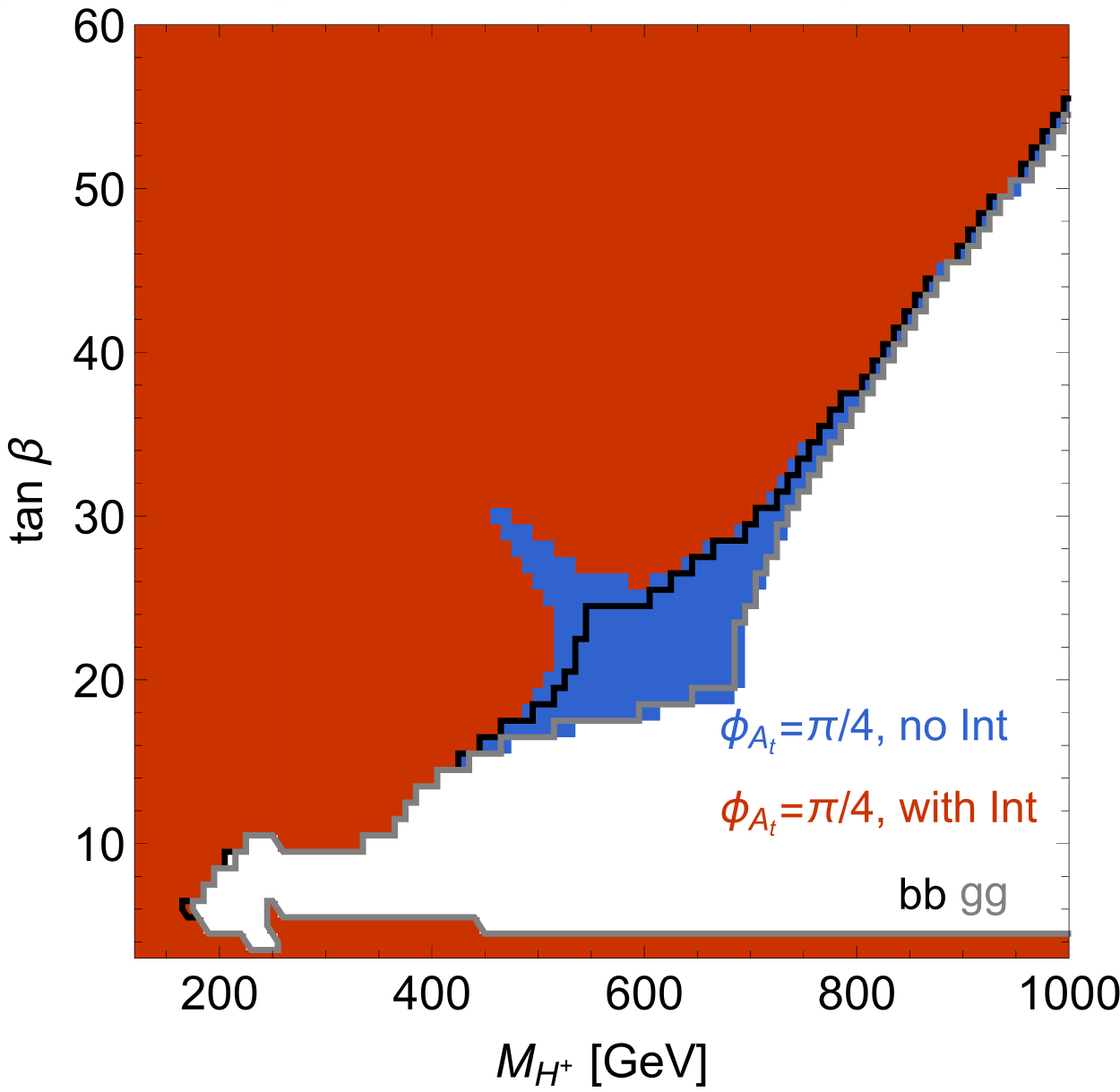}
  \caption{Exclusion bounds obtained with \texttt{HiggsBounds} in the complex $\mhmod$-scenario with $\mu=1000\gev$ and $\pat=\pi/4$: 
  without the interference term (blue), including the interference term in the $\bb$ \textit{and} the $gg$ processes (red), including the interference \textit{only} in $\bb$ (black line) or \textit{only} in $gg$ (grey line).
  }
  \label{fig:HB}
 \end{center}
\end{figure}

Fig.\,\ref{fig:HB} shows the resulting exclusion bounds obtained with \texttt{HiggsBounds-4.2.0} (including Higgs search results up to run 1 of the LHC) with MSSM input from \texttt{FeynHiggs}, in the complex $\mhmod$-scenario with $\mu=1000\gev$ and $\pat=\frac{\pi}{4}$. The augmented value of $\mu$ with respect to the default value of $200\gev$ closes the decay channel into higgsinos and thereby increases ${\rm BR}(h_a\rightarrow\tau\tau)$. Besides, it enhances the $\cp$-violating effects caused by the product $\mu A_t$ where $A_t$ has an imaginary part.
The blue area represents the parameter space that seems to be excluded at the 95\% confidence level (CL) in this scenario if the interference term is neglected by employing the standard NWA despite the presence of the complex phase $\pat$. 
This exclusion limit based on the CP-violating $3\times 3$ mixing of $h_1, h_2, h_3$ is stronger than in the case of real parameters (with only CP-conserving $2\times 2$ mixing) because the large off-diagonal elements in the full $\Zb$-mixing matrix enhance each individual production cross section $\sigma(P\rightarrow h_a)$ for $a=2,3$ in the region where $h_2$ and $h_3$ are quasi degenerate.
The grey line corresponds to the case where the interference term is only included in the $gg$-initiated process whereas the inclusion of the interference term only in the $\bb$ process gives rise to the black line.
In contrast, the interference term is taken into account in both production processes in the 95\% CL exclusion contour displayed in red. The direct comparison shows that a substantial area between $\mhp\simeq 450\gev$ and $700\gev$ and roughly from $\tan\beta\simeq 18$ to $32$ cannot be excluded in this scenario with complex parameters due to the strongly destructive interference effect observed in Fig.\,\ref{fig:eta} in the same parameter region. This discrepancy between the conventional approach of the incoherent sum of separate Higgs contributions and the approach of including the interference among different neutral Higgs bosons shows that the standard NWA is insufficient in such a $\cp$-violating scenario. Interference between $h$ and $H$ can also play a role in the MSSM with real parameters, but it becomes relevant only in a narrow parameter region at low $\mhp$ and very high $\tan\beta$, which lies in the deeply excluded region. However, the mass difference between $M_{h_2}$ and $M_{h_3}$ is smaller than $\Gamma_{h_2}+\Gamma_{h_3}$ in the major part of the parameter plane so that the $h_2-h_3$ interference term becomes important 
in the case of $\cp$-violating mixing. 

\section{Conclusions}
$CP$-violating mixing and interference of neutral Higgs bosons in the MSSM with complex parameters can have a significant impact on the interpretation of LHC searches for additional scalars. We have investigated a $\cp$-violating benchmark scenario with a substantial mixing of $H$ and $A$ into the quasi-degenerate states $h_2$ and $h_3$. Both in $b\bar{b}$-associated Higgs production and in gluon fusion of $h_1,h_2,h_3$ decaying into a pair of $\tau$-leptons, a very large, destructive interference effect is found. The individual single-particle cross section parts are enhanced by mixing effects whereas the combined cross section is substantially suppressed by the interference. As a consequence, a considerable parameter region, which would seem to be ruled out if the interference term were neglected, actually escapes exclusion by LHC Run 1 data.
We will extend\,\cite{IntCalc:InProgress} our analysis to the recent Run 2 results, additional phases of complex parameters and Higgs production cross sections from \texttt{SusHiMi}\,\cite{Liebler:2016ceh} that take the full phase dependence into account.

\section*{Acknowledgements}

We would like to thank Oscar St\aa l for his help with \texttt{HiggsBounds} and valuable suggestions, and Christian Veelken, Sven Heinemeyer, Stefan Liebler and Shruti Patel for useful discussions. EF thanks the DESY theory group for hospitality. The work of EF was partially funded by the German National Academic Foundation and by DESY. The work of GW\ is supported in part by the Collaborative Research Centre SFB~676 of the DFG, ``Particles, Strings and 
the Early Universe'', and by the European Commission through the ``HiggsTools'' 
Initial Training Network PITN-GA-2012-316704.

\bibliographystyle{unsrt}

\begin{thebibliography}{99}
\bibitem{Fuchs:2014ola}
  E.~Fuchs, S.~Thewes and G.~Weiglein,
  Eur.\ Phys.\ J.\ C {\bf 75} (2015) 254.
 
  
\bibitem{Fuchs:2015jwa}
  E.~Fuchs,
  DESY-THESIS-2015-037.
  
\bibitem{Fuchs:2017wkq}
  E.~Fuchs and G.~Weiglein,
  arXiv:1705.05757 [hep-ph].
  
\bibitem{Fuchs:2016swt}
  E.~Fuchs and G.~Weiglein,
  arXiv:1610.06193 [hep-ph].

  
\bibitem{Carena:2013ytb}
  M.~Carena, S.~Heinemeyer, O.~St\aa l, C.~E.~M.~Wagner and G.~Weiglein,
  Eur.\ Phys.\ J.\ C {\bf 73} (2013) 9,  2552.


  
  
\bibitem{Heinemeyer:1998np}
  S.~Heinemeyer, W.~Hollik and G.~Weiglein,
  Eur.\ Phys.\ J.\ C {\bf 9} (1999) 343.

\bibitem{Heinemeyer:1998yj}
  S.~Heinemeyer, W.~Hollik and G.~Weiglein,
  Comput.\ Phys.\ Commun.\  {\bf 124} (2000) 76.
  [hep-ph/9812320].

\bibitem{Degrassi:2002fi}
  G.~Degrassi, S.~Heinemeyer, W.~Hollik, P.~Slavich and G.~Weiglein,
  Eur.\ Phys.\ J.\ C {\bf 28} (2003) 133.

\bibitem{Heinemeyer:2007aq}
  S.~Heinemeyer, W.~Hollik, H.~Rzehak and G.~Weiglein,
  Phys.\ Lett.\ B {\bf 652} (2007) 300.

\bibitem{Hahn:2013ria}
  T.~Hahn, S.~Heinemeyer, W.~Hollik, H.~Rzehak and G.~Weiglein,
  Phys.\ Rev.\ Lett.\  {\bf 112} (2014) 14,  141801.

\bibitem{Frank:2006yh}
  M.~Frank, T.~Hahn, S.~Heinemeyer, W.~Hollik, H.~Rzehak and G.~Weiglein,
  JHEP {\bf 0702} (2007) 047.





\bibitem{Barger:2001nu}
  V.~D.~Barger, T.~Falk, T.~Han, J.~Jiang, T.~Li and T.~Plehn,
  Phys.\ Rev.\ D {\bf 64} (2001) 056007.

\bibitem{Li:2010ax}
  Y.~Li, S.~Profumo and M.~Ramsey-Musolf,
  JHEP {\bf 1008} (2010) 062.

\bibitem{Engel:2013lsa}
  J.~Engel, M.~J.~Ramsey-Musolf and U.~van Kolck,
  Prog.\ Part.\ Nucl.\ Phys.\  {\bf 71} (2013) 21.

\bibitem{IntCalc:InProgress}
      E.~Fuchs, S.~Liebler, S.~Patel and G.~Weiglein,
      in preparation.

%


\bibitem{CMS:2015mca}
  CMS Collaboration [CMS Collaboration],
  CMS-PAS-HIG-14-029.
  
\bibitem{Aad:2014vgg}
  G.~Aad {\it et al.} [ATLAS Collaboration],
  JHEP {\bf 1411} (2014) 056.

\bibitem{CMS:2016pkt}
  CMS Collaboration [CMS Collaboration],
  CMS-PAS-HIG-16-006.
  
\bibitem{Aaboud:2016cre}
  M.~Aaboud {\it et al.} [ATLAS Collaboration],
  Eur.\ Phys.\ J.\ C {\bf 76} (2016) no.11,  585



  
\bibitem{Bernreuther:2015fts}
  W.~Bernreuther, P.~Galler, C.~Mellein, Z.~G.~Si and P.~Uwer,
  Phys.\ Rev.\ D {\bf 93} (2016) no.3,  034032
  
\bibitem{Carena:2016npr}
  M.~Carena and Z.~Liu,
  JHEP {\bf 1611} (2016) 159

  
\bibitem{Bechtle:2008jh}  
  P.~Bechtle, O.~Brein, S.~Heinemeyer, G.~Weiglein and K.~E.~Williams,
  Comput.\ Phys.\ Commun.\  {\bf 181} (2010) 138.
  
\bibitem{Bechtle:2011sb}
  P.~Bechtle, O.~Brein, S.~Heinemeyer, G.~Weiglein and K.~E.~Williams,
  Comput.\ Phys.\ Commun.\  {\bf 182} (2011) 2605.
  
\bibitem{Bechtle:2013gu}
  P.~Bechtle, O.~Brein, S.~Heinemeyer, O.~St\aa l, T.~Stefaniak, G.~Weiglein and K.~Williams,
  PoS CHARGED {\bf 2012} (2012) 024.
  
\bibitem{Bechtle:2013wla}
  P.~Bechtle, O.~Brein, S.~Heinemeyer, O.~St\aa l, T.~Stefaniak, G.~Weiglein and K.~E.~Williams,
  Eur.\ Phys.\ J.\ C {\bf 74} (2014) 3,  2693.
  
\bibitem{Liebler:2016ceh}
  S.~Liebler, S.~Patel and G.~Weiglein,
  Eur.\ Phys.\ J.\ C {\bf 77} (2017) no.5,  305
 

\end{thebibliography}

\end{document}